\documentclass[aps, superscriptaddress, showpacs]{revtex4}
\usepackage{epsfig}
\usepackage{graphicx}
\usepackage{amsmath}
\usepackage{bm}
\usepackage{amssymb}
\usepackage{color}

\newcommand{\be}{\begin{equation}}
\newcommand{\ee}{\end{equation}}
\newcommand{\ba}{\begin{eqnarray}}
\newcommand{\ea}{\end{eqnarray}}

\usepackage[normalem]{ulem}

\begin{document}

\title{Coulomb center instability in bilayer graphene}
\date{\today}

\author{D. O. Oriekhov}
\affiliation{Department of Physics, Taras Shevchenko National University of Kiev, Kiev, 03680, Ukraine}

\author{O. O. Sobol}
\affiliation{Department of Physics, Taras Shevchenko National University of Kiev, Kiev, 03680, Ukraine}

\author{E. V. Gorbar}
\affiliation{Department of Physics, Taras Shevchenko National University of Kiev, Kiev, 03680, Ukraine}
\affiliation{Bogolyubov Institute for Theoretical Physics, Kiev, 03680, Ukraine}

\author{V. P. Gusynin}
\affiliation{Bogolyubov Institute for Theoretical Physics, Kiev, 03680, Ukraine}

\begin{abstract}
In the low-energy two-band as well as four-band continuum models, we study the supercritical charge instability in gapped bilayer graphene in the field of an impurity charge when the lowest-energy bound
state dives into the hole continuum. It is found that the screening effects are crucially important in bilayer graphene. If they are neglected, then the critical value for the impurity charge  tends to zero
as the gap $\Delta$ vanishes. If the screened Coulomb interaction is considered, then the critical charge tends to a finite value for $\Delta \to 0$. The different scalings of the kinetic energy of quasiparticles and the Coulomb interaction with respect to the distance to the charged impurity ensure that the wave function of the electron bound state does not shrink toward the impurity as its charge increases. This results in the absence of the fall-to-center phenomenon in bilayer graphene although the supercritical charge instability is realized.
\end{abstract}
\maketitle

\section{Introduction}

For a long time \cite{Pomeranchuk}, the problem of the atomic collapse in relativistic quantum mechanical systems has attracted significant interest. Historically the first and perhaps most important example of the phenomenon of the fall to center is connected with the supercritical instability of
electrons in the Coulomb field of heavy nuclei \cite{Zeldovich,Greiner}. This instability occurs when
the electron bound state of the lowest energy traverses the energy distance separating the upper and
lower continua as the charge of the nucleus increases and dives into the lower continuum producing a resonance, which describes an emitted positron. The corresponding critical charge is estimated to be
of order $Z_{cr}\simeq170$ \cite{Zeldovich,Greiner}. Unfortunately, it is difficult to observe the
phenomenon of the atomic collapse in QED because nuclei with charge $Z \ge 170$ are not encountered in nature.

In single-layer graphene, it was found that its electron excitations at low energies are described
by the massless Dirac equation in two spatial dimensions (2D) \cite{Semenoff}. These excitations in
the field of a charged impurity realize a 2D analog of the problem of the atomic collapse in QED. The theoretical studies performed in Refs.~[\onlinecite{Pereira,Shytov,Novikov}] showed that the
fall to center does take place in gapless graphene if the Coulomb potential strength exceeds a certain critical value $Z_{cr}\alpha_g/\kappa=1/2$ where $\alpha_g=e^2/\hbar v_F\sim1$ and $\kappa$ is the dielectric constant. This critical value $Z_{cr}$ is dramatically smaller than that in QED. The main
reason for such a reduction is the large effective coupling constant $\alpha_g$ in graphene, which  is approximately 300 times larger than the QED coupling constant $\alpha$ due to the Fermi velocity $v_F \approx 10^{6}\,{\rm m/s}$. In gapped graphene $Z_{cr}\alpha_g/\kappa$ becomes a function of a gap $\Delta$
and $Z_{cr}\alpha_g/\kappa\to 1/2$ when $\Delta$ goes to zero \cite{excitonic-instability}. The supercritical Coulomb center instability is also closely related to the excitonic instability in graphene
in the strong-coupling regime $\alpha_g > \alpha_c\sim1$ (see Refs.\cite{excitonic-instability,Fertig,Guinea}) and to the gap opening, which may transform graphene into an insulator \cite{metal-insulator,GGG2010,MS-phase-transition,Wang-Liu,Gonzalez,Popovici}. Indeed, the electron in the strong-coupling regime $\alpha_g>\alpha_{c}$ can spontaneously create from the vacuum electron-hole pairs (in the same way as the supercritical charge creates electron-hole pairs). While the initial electron attracts the hole and forms the bound state (exciton), the emitted electron with supercritical charge can spontaneously create another electron-hole pair. Therefore, the process of
creating pairs continues leading to the formation of the exciton condensate and, as a result, the quasiparticles acquire a gap.

Experimentally, the supercritical resonances were observed \cite{Wang} in graphene by collecting
clusters of Ca dimers confirming, thus, for the first time the phenomenon of the supercritical
instability in a system with a relativistic-like energy spectrum. Also, the electron states in the field
of a Coulomb impurity and in the presence of a magnetic field were studied experimentally and
theoretically in Refs.\cite{center-in-mag.field,Mao,Sobol2016}, where it was shown that the strength
of a charged impurity can be effectively tuned by controlling the occupation of Landau-level states
with a gate voltage.

Bernal-stacked bilayer graphene forms another very interesting two-dimensional physical system
whose electron excitations are described by the Hamiltonian with nontrivial chiral properties \cite{bilayer,Novoselov}. These chiral fermions do not have  analogs in high-energy physics because
their energy spectrum $E \sim \pm|\mathbf{p}|^2$ is parabolic in the two-band model, like in nonrelativistic systems described by the Schr\"odinger equation. Still since there are positive and negative energy bands related through the charge conjugation, this suggests that the instability
connected with diving of the lowest-energy electron bound state into the lower band may take place in
gapped bilayer graphene too. The combined effects of chirality and the
parabolic dispersion of quasiparticles in gapless bilayer graphene result, as was shown in Ref.~[\onlinecite{Shytov-states}], in cloaked bound states in the $p$-wave channel that do not hybridize with the hole continuum. This ensures that the electrons occupying the lower continuum cannot populate the cloaked states, 
and as a result, the supercritical instability does not take place due to these states. In Sec.~\ref{four-band-model} we will show that in contrast to Ref.~[\onlinecite{Shytov-states}] the
$s$-wave bound states in gapped bilayer graphene hybridize with the lower continuum states leading to
the supercritical instability connected with the emission of electron-hole pairs.

Actually, there is more to this. The parabolic spectrum in bilayer graphene is more soft than in
monolayer graphene. Therefore, the electron-electron interactions are effectively enhanced and
may open a gap connected with the condensation of electron-hole pairs in bilayer graphene. This
conclusion agrees with the experimental data. While no gap is observed in monolayer graphene at
the neutrality point, a small gap 2 $\mbox{meV}$ is realized in bilayer graphene
\cite{Martin,Weitz,Freitag} in the absence of external electromagnetic fields. A much larger gap
($\sim42$ meV) is observed in high-mobility ABC-stacked trilayer graphene \cite{trilayer} where
electron excitations have a softer dispersion $E(\mathbf{p})\sim \pm|\mathbf{p}|^3$.
As we mentioned above, the supercritical instability in the Coulomb center is a precursor of
the excitonic instability and gap opening in the electron spectrum of a many-body problem.
Reversing this argument and taking into account the fact that bilayer graphene is gapped due to
the electron-electron interactions, one may expect that the supercritical instability for the
electron in the field of the Coulomb center should take place in bilayer graphene.

The following qualitative consideration is important for the analysis of the supercritical instability
in bilayer graphene. Since the quasiparticle kinetic energy scales like $1/r^2$ with distance from
the charged impurity in bilayer graphene and is negligible compared to the Coulomb interaction $1/r$,
this suggests the absence of the critical charge for the formation of a bound state in this material.
On the other hand, since the kinetic energy at small distances is larger than the Coulomb interaction energy, this implies that the fall to center should not take place in bilayer graphene. Therefore,
{\it a priori}, one may expect that the supercritical instability in bilayer graphene should not be necessarily related to the phenomenon of the fall to center and, therefore, these two phenomena, in
general, are distinct. It is appropriate to discuss the nomenclature that we use in this paper. By the fall-to-center phenomenon we understand the situation when the wave function of the lowest-energy bound
state is localized in a region which shrinks to the charged impurity as the parameter regularizing the Coulomb interaction at small distances is removed. The supercritical instability and atomic collapse are synonymous and take place when the lowest energy bound state of the charged impurity problem in bilayer graphene dives into the lower continuum making it unstable with respect to the emission of the electron-hole pairs. The situation is different in monolayer graphene where the supercritical instability is directly related to the fall-to-center phenomenon. The physical reason for this is the equal scaling of the kinetic energy and Coulomb interaction with distance.

Clearly, the above heuristic reasoning is based on the scaling in the effective two-band low-energy model of bilayer graphene. However, the parabolic spectrum in bilayer graphene \cite{bilayer} is valid only up
to momenta $|\mathbf{p}| \approx \gamma_1/(2v_F)$, where $\gamma_1=0.39\,\mbox{eV}$ is the interlayer hopping amplitude. For large momenta, the energy dispersion is linear as in monolayer graphene. Therefore, it is possible, in principle, that as the charge of the impurity increases the wave function
of the electron bound state becomes more localized in the vicinity of the impurity such that the
parabolic energy spectrum does not apply. If this happens, then the supercritical instability in bilayer graphene would proceed like in monolayer graphene and would be connected with the fall-to-center phenomenon.

In order to see whether the supercritical instability is related to the fall to center in
bilayer graphene, we study in this paper the electron bound states in the field of a charged impurity
in bilayer graphene by using the two continuum models: the low-energy two-band model as well as the four-band model. The main principal advantage of the four-band model for the study of supercritical instability in bilayer graphene is that it takes into account the evolution of the electron energy dispersion from the low-energy quadratic to high-energy linear energy dispersion. Since the trigonal
warping is small in bilayer graphene, we will neglect it in  our study. On the other hand, the screening effects play a very essential role in view of the finite density of states at zero energy in bilayer graphene and will be taken into account in our study. We will show that while the lowest-energy bound
state in the Coulomb center in gapped bilayer graphene indeed dives into the lower continuum, the wave function of the electron bound state does not shrink toward the impurity as its charge increases. This
means that the supercritical instability in bilayer graphene is not related to the phenomenon of the fall to center.

The paper is organized as follows. In Sec.\ref{section-model}, we consider the electron states in the Coulomb field of a charged impurity in bilayer graphene in the low-energy two-band model. The variational method is employed in Sec.\ref{section-variational} in order to study the supercritical instability in
the cases of the bare and screened Coulomb interactions. Numerical results of our analysis of the supercritical instability in the two-band model are given in Sec.\ref{section-numerical}. The supercritical instability in the four-band model is analyzed in Sec.\ref{four-band-model}. The results are summarized
and discussed in Sec.\ref{section-conclusion}.

\section{Two-band model}
\label{section-model}

The two-band Hamiltonian, which describes the low-energy electron excitations in gapped bilayer graphene
in the field of an impurity with charge $Z e$, reads \cite{bilayer}
  \begin{equation}
  H=\frac{v^2_F}{\gamma_1}\left(\begin{array}{cc}
  0&(p_{-})^2\\
  (p_{+})^2&0
  \end{array}\right)+\Delta\left(\begin{array}{cc}
  1&0\\
  0&-1
  \end{array}\right)+V(r),\quad V(r)=-\frac{Ze^{2}}{\kappa\,r},
  \label{Hamiltonian}
  \end{equation}
where $p_\pm=p_x\pm i p_y$ and $\bf{p}=-i\hbar\bm{\nabla}$ is the two-dimensional momentum operator,
$\gamma_1 \approx 0.39~\mbox{eV}$ is the strongest interlayer coupling between the pairs of orbitals that
lie directly below and above each other, and we choose the dielectric constant $\kappa=4$ in our numerical calculations in this paper.

The energy spectrum of the free part of the Hamiltonian (\ref{Hamiltonian}) is given by
$E({\bf p})=\pm\sqrt{({\bf p}^2/2m^{*})^2+\Delta^2}$, so that we have a gap $2\Delta$ between the lower
and upper continua (valence and conductivity bands), where $m^{*}=\gamma_1/2v_F^2\approx0.054m_e$ is
the quasiparticle mass and $m_e$ is the mass of the electron. Although the kinetic part of Hamiltonian (\ref{Hamiltonian}) has the same matrix structure as in monolayer graphene leading to chirality and the particle-hole symmetry characteristic of relativistic systems, the dispersion relation for quasiparticles
in bilayer graphene is quadratic like in nonrelativistic systems. Thus, the low-energy Hamiltonian (\ref{Hamiltonian}) nontrivially combines relativistic and nonrelativistic properties. Concerning
the form of the Coulomb potential, we note that the charged impurity is usually displaced away from the
graphene plane by some distance $d$ or is smeared over some region that leads to a regularized form of
the Coulomb interaction $V(r) \sim 1/\sqrt{r^2+d^2}$ \cite{Novikov}. While the regularization of the Coulomb potential is crucial for the study of a charge instability in monolayer graphene, it is sufficient to
use the standard potential $\sim 1/r$ for the two-band model of bilayer graphene with the parabolic dispersion of quasiparticles.

As we discussed in the Introduction, there is a certain critical value of the impurity charge in monolayer graphene above which the atomic collapse occurs. It is interesting to see what happens in bilayer graphene in view of the different quadratic dispersion relation in this material. Although the two-band model is applicable only up to the energies of order $\gamma_1/4$ when the next band becomes important, it still makes sense to study how the supercritical Coulomb center instability is realized in the low-energy effective two-band model. We perform such a study in this section. The four-band model will be analyzed
in Sec. \ref{four-band-model}, where we will see that the two- and four-band models give very close results.

It is convenient to define the coupling constant as $\xi=\frac{Z\alpha_g}{\kappa}\sqrt{\frac{\gamma_{1}}{\Delta}}$ and express distances in terms of a characteristic length $\lambda_{\Delta}={\hbar v_{F}}/{\sqrt{\gamma_{1}\Delta}}$ and wave vectors in its inverse, i.e., $\mathbf{\tilde{k}}=\mathbf{k}\,\lambda_{\Delta}$ (in what follows we will omit the tilde over dimensionless momenta). The eigenstates of Hamiltonian (\ref{Hamiltonian}) for $\Psi^T=(\chi,\varphi)$ are determined in momentum space by the following system of equations:
  \begin{equation}
  \label{35}
  \begin{array}{c}
  {(k_{x}-ik_{y})^2\varphi\left(\mathbf{k}\right)}+{(1-\epsilon)\chi\left(\mathbf{k}\right)}+
  {\int\!\!\frac{d^2q}{(2\pi)^{2}}\,{{\chi{\left(\mathbf{q}\right)}}V_{\rm eff}\left(\mathbf{k}-\mathbf{q}\right)}} = 0,\\
  {(k_{x}+ik_{y})^2\chi\left(\mathbf{k}\right)}-{(1+\epsilon)\varphi\left(\mathbf{k}\right)}
  +{\int\!\!\frac{d^2q}{(2\pi)^{2}}\,{{\varphi{\left(\mathbf{q}\right)}}V_{\rm eff}
  \left(\mathbf{k}-\mathbf{q}\right)}} = 0,
  \end{array}
  \end{equation}
where $\epsilon=E/\Delta$ and $V_{\rm eff}$ describes the screened potential of the impurity of charge $Ze$.
It is defined by an analog of the Poisson equation of the form
  \begin{equation}
  \label{Poisson_eq}
  \sqrt{-\Delta_{2D}}V_{\rm eff}(\mathbf{x})=-2\pi\xi \delta^{(2)}(\mathbf{x})
  -\frac{4\pi\alpha_g}{\kappa}\int\!\!d^{2}\mathbf{y}\,\Pi(\mathbf{x}-\mathbf{y})V_{\rm eff}(\mathbf{y}),
  \end{equation}
where $\Pi(\mathbf{x})$ is the static polarization function in bilayer graphene. Notice the presence
of the pseudodifferential operator $\sqrt{-\Delta_{2D}}$ on the left-hand side of the equation above,
which is necessary in order to correctly describe the Coulomb interaction in a dimensionally reduced electrodynamical system \cite{reduced}. In momentum space, we easily find the following solution to Eq.~(\ref{Poisson_eq}):
  \begin{equation}
  \label{Vtot1}
  V_{\rm eff}(q)=-\frac{2\pi\xi}{q+\frac{4\pi\alpha_g}{\kappa}\Pi(q)} \equiv -2\pi\xi
  \left(\frac{1}{q}-\delta V(q)\right),
  \end{equation}
where
\begin{equation}
\delta V(q)=\frac{4\pi\alpha_g}{\kappa}\frac{\Pi(q)}{q\left[q+\frac{4\pi\alpha_g}{\kappa}\Pi(q)\right]}
\label{correction-interaction}
\end{equation}
is the correction to the Coulomb interaction due to the screening effects. The one-loop polarization function as an integral over momentum was derived in Ref.~[\onlinecite{Levitov}] and equals
\begin{equation}
  	 \Pi(\mathbf{q})=\sqrt{\frac{\gamma_1}{\Delta}}\int \frac{d^2 k}{(2\pi)^2} \frac{2(\epsilon_{-}\epsilon_{+}-1)-(\mathbf{k}
  	 -\frac{\mathbf{q}}{2})_{+}^{2}(\mathbf{k}
  	 +\frac{\mathbf{q}}{2})_{-}^{2}-(\mathbf{k}+\frac{\mathbf{q}}{2})_{+}^{2}(\mathbf{k}
  	 -\frac{\mathbf{q}}{2})_{-}^{2}}{\epsilon_{-}\epsilon_{+}(\epsilon_{-}
  	 +\epsilon_{+})},\quad   \epsilon_{\pm}=\sqrt{1+\left|\mathbf{k}\pm\frac{\mathbf{q}}{2}\right|^{4}}.
  	\label{exact_polariz}
  \end{equation}
For $q \ll 1$, the polarization function is proportional to $q^{2}$ and tends to zero for $q\to 0$; therefore, $\delta V(q)$ is not singular at $q=0$. For $q \gg 1$, $\Pi(q)=\sqrt{\gamma_{1}/\Delta}\ln2/\pi$.

In polar coordinates, the angle and the absolute value of momentum variables can be separated.
In order to do this, we note that Hamiltonian (\ref{Hamiltonian}) does not conserve the $z$ component of
the pseudospin $S_z=\hbar\sigma_z$ and the angular momentum $L_z=xp_y-yp_x=-i\hbar\partial/\partial\theta$ separately because they do not commute with the Hamiltonian. However, the $z$ component of the total
angular momentum commutes with the Hamiltonian
\begin{equation}
J_z=L_z+S_z=-i\hbar\frac{\partial}{\partial\theta}+\hbar\sigma_z,\quad\quad [J_z,\hat{H}]=0.
\end{equation}
Thus, we seek the spinor function in the form
\begin{equation}
  \label{spinor}
  \Psi=\left(\begin{array}{c}
  \chi(\mathbf{k})\\
  \varphi(\mathbf{k})
  \end{array}\right)=\left(\begin{array}{c}
  a_j(k)e^{i(j-1)\theta}\\
  b_j(k)e^{i(j+1)\theta}
  \end{array}\right),
\end{equation}
where $j$ is the total angular momentum which takes integer values $j=0,\pm1,\pm2,\dots$. Taking into account $|\mathbf{k}-\mathbf{q}|=\sqrt{k^2+q^2-2kq\cos(\theta-\phi)}$ and $(k_{x}\pm ik_{y})^2=k^2e^{\pm 2i\theta}$, the system of equations (\ref{35}) takes the form
  \begin{equation}
  \label{39}
  \begin{array}{c}
  k^{2}b_{j}(k) + a_{j}(k) - {\frac{\xi}{2\pi}}{\int\limits_{0}^{\Lambda}dq q a_{j}(q)  \int\limits_{0}^{2\pi}d\phi\left[\frac{1}{|\mathbf{k}-\mathbf{q}|}
  -\delta V(\mathbf{k}-\mathbf{q})\right]e^{i(j-1)(\phi-\theta)}} = \epsilon a_j(k),\\
  k^{2}a_j(k)-b_j(k)-{\frac{\xi}{2\pi}}{\int\limits_{0}^{\Lambda}dq q b_{j}(q) \int\limits_{0}^{2\pi}d\phi\left[\frac{1}{|\mathbf{k}-\mathbf{q}|}
  -\delta V(\mathbf{k}-\mathbf{q})\right]e^{i(j+1)(\phi-\theta)}} = \epsilon b_{j}(k),
  \end{array}
  \end{equation}
where $\Lambda=\sqrt{\gamma_{1}/(4\Delta)}$ is the $UV$ cutoff of the low-energy effective Hamiltonian (\ref{Hamiltonian}).

Integrating over angle, we have
  \begin{equation}
  K_{j}(k, q)=\int\limits_{0}^{2\pi} \frac{d\phi\cos(j\phi)}{\sqrt{k^2+q^2-2kq \cos\phi}}  = \frac{2}{\sqrt{kq}}Q_{|j|-1/2}\left(\frac{k^2+q^2}{2kq}\right),
\label{kernel}
  \end{equation}
where $Q_\nu(z)$ is the Legendre function of the second kind. Denoting
  \begin{equation}
  \delta V_{j}(k, q)=\int\limits_{0}^{2\pi}d\phi \,\delta V\left(\sqrt{k^2+q^2-2kq \cos\phi}\right)\,\cos(j\phi),
  \end{equation}
we find that the system of equations (\ref{39}) takes the form
  \begin{equation}
  \label{system}
  \begin{array}{c}
  k^{2}b_j(k) + a_j(k) - {\frac{\xi}{2\pi}}{\int\limits_{0}^{\Lambda}dq q\,a_j(q) [K_{j-1}(k,q)-
  \delta V_{j-1}(k,q)]} = \epsilon a_j(k),\\
  k^{2}a_j(k)-b_j(k)-{\frac{\xi}{2\pi}}{\int\limits_{0}^{\Lambda}dq q\,b_j(q) [K_{j+1}(k,q)-
  \delta V_{j+1}(k,q)]} = \epsilon b_j(k),
  \end{array}
  \end{equation}
where the kernels $K_{j}(k,q)$ can be expressed in terms of the full elliptic integrals of the first
and second kind (see Appendix). Unfortunately, it is not possible to find analytically a solution to the above system of equations. Its numerical solutions will be given in Sec.\ref{section-numerical}. Still,
in order to have an analytic insight into the problem under consideration, we will apply to it the variational method in the next section. A very significant advantage of the variational method is its transparency and the possibility of analytic control.

\section{Variational method}
\label{section-variational}

It is well known that the variational method is very efficient if the asymptotes of the wave function are
known. In order to find them, we seek eigenstates of Hamiltonian (\ref{Hamiltonian})
\begin{equation}
\label{substitution}
\Psi_j(r)=\left(\begin{array}{c}e^{i(j-1)\theta}a_j(r)\\e^{i(j+1) \theta}b_j(r)\end{array}\right)
\end{equation}
as eigenstates of $J_z$ too,
\begin{equation}
J_z\Psi_j(r)=j\Psi_j(r).
\end{equation}
The stationary Schr$\ddot{o}$dinger equation for quasiparticles in bilayer graphene in the field of
a spherically symmetric potential $V(r)$ defines the following system of equations for the components
of spinor (\ref{substitution}):
\begin{equation}
 \begin{array}{c}
 \frac{\hbar^2}{2m^{*}} \left(\frac{d^2}{d\,r^2}-\frac{2\,j-1}{r}{\frac{d}{d\,r}}+\frac{j^{2}-1}{r^2}\right)a_j(r) = -(\Delta+E-V(r))b_j(r),\\
\frac{\hbar^2}{2m^{*}} \left({\frac{d^2}{d\,r^2}}+{\frac{2\,j+1}{r}{\frac{d}{d\,r}}}+{\frac{j^{2}-1}{r^2}}\right)b_j(r) = (\Delta-E+V(r))a_j(r).
\end{array}
\label{configspace:eq}
 \end{equation}
It is convenient to use the functions $f(r)=\sqrt{r}\,a_j(r)$ and $g(r)=\sqrt{r}\,b_j(r)$ and express
$g(r)$ through $f(r)$ by using the first equation of the above system:
\begin{equation}
g(r)=-\frac{{\hbar^2}/{2m^*}}{\Delta+E-V(r)}\left(\frac{d^2}{dr^2}-\frac{2j}{r}
\frac{d}{dr}+\frac{(j+1/2)^2-1}{r^2}\right)f(r).
\end{equation}
Substituting it in the second equation, we obtain the equation for $f(r)$:
\begin{eqnarray}
\left(\frac{d^2}{dr^2}+\frac{2j}{r}\frac{d}{dr}+\frac{(j-1/2)^2-1}{r^2}\right)
\left[\frac{({\hbar^2}/{2m^*})^2}{\Delta+E-V(r)}\left(\frac{d^2}{dr^2}
-\frac{2j}{r}\frac{d}{dr}+\frac{(j+1/2)^2-1}{r^2}\right)\right]f(r)
=(E-\Delta-V(r))f(r).
\label{eq:f(r)-4thorder}
\end{eqnarray}
Multiplying Eq.~(\ref{eq:f(r)-4thorder}) by $f^*(r)$ and integrating with respect to $r$, we find the following equation:
\begin{equation}
\int\limits_0^\infty dr\frac{({\hbar^2}/{2m^*})^2|L(j)f|^2}{E+\Delta-V(r)}=\int\limits_0^\infty dr(E-\Delta-V(r))|f(r)|^2,\quad L(j)
=\frac{d^2}{dr^2}-\frac{2j}{r}\frac{d}{dr}+
\frac{(j+1/2)^2-1}{r^2},
\label{functional_f-4th-order}
\end{equation}
which is defined for any values of $E$ in the interval $[-\Delta,\Delta]$ when $V(r)$ is negative. Obviously, Eq.~(\ref{functional_f-4th-order}) is exactly satisfied for the solutions to Eq.~(\ref{eq:f(r)-4thorder}). For an arbitrary function $f(r)$, Eq.~(\ref{functional_f-4th-order}) can
be considered as a nonlinear functional $f\mapsto E[f]$. For a given function $f$, while the left-hand
side of the above equation decreases with $E$, its right-hand side increases for $E$ in the interval $[-\Delta,\Delta]$. Therefore, if we choose an ansatz for $f$ with some variational parameters for which
Eq.~(\ref{functional_f-4th-order}) is satisfied, then this functional defines energy $E$ as a
unique function of the variational parameters. According to Refs.~\cite{Dolbeault1,Dolbeault2}, the
ground bound state  corresponding to the lowest energy $E_g$ is realized, as usual in the variational
method, as the global minimum of the bounded from below functional $E[f]$, i.e.,
\begin{equation}
E_g={\rm min}_f\,E[f].
\end{equation}
We will apply this variational method in the next two subsections in order to find the critical charge
when the lowest-energy bound state reaches the upper boundary $E=-\Delta$ of the negative continuum. We
will consider the cases of the bare and screened Coulomb interaction.

\subsection{Coulomb potential}

Let us begin our variational analysis with the case of the unscreened Coulomb interaction $V_C(r)=-Ze^2/(\kappa\,r)$. For this interaction, Eq.(\ref{eq:f(r)-4thorder}) in terms of $\lambda_\Delta=\hbar v_F/\sqrt{\gamma_1\Delta}$ takes the form
\begin{eqnarray}
&&\left(\frac{d^2}{dr^2}+\frac{2j}{r}\frac{d}{dr}+\frac{(j-1/2)^2-1}{r^2}\right)
\left[\frac{1}{\frac{E}{\Delta}+1+\frac{\xi}{r}}\left(\frac{d^2}{dr^2}-\frac{2j}{r}
\frac{d}{dr}\right.\right.\nonumber\\
&&\left.\left.+\frac{(j+1/2)^2-1}{r^2}\right)\right]f(r)=(\frac{E}{\Delta}-1+\frac{\xi}{r})f(r),\quad
\xi=\frac{Z\alpha_g}{\kappa}\sqrt{\frac{\gamma_1}{\Delta}}.
\label{eq:f(r)-4order-dimless}
\end{eqnarray}
In order to find the critical charge, we should set $E=-\Delta$ and study the equation for the
state with $j=1$ which defines the bound state of the lowest energy:
\begin{equation}
\left(\frac{d^2}{dr^2}+\frac{2}{r}\frac{d}{dr}-\frac{3}{4r^2}\right)
\left[r\left(\frac{d^2}{dr^2}-\frac{2}{r}\frac{d}{dr}+\frac{5}{4r^2}\right)\right]
f(r)=-\xi\left(2-\frac{\xi}{r}\right)f(r),
\label{eq:f-critical}
\end{equation}
or in the equivalent form,
\begin{equation}
L_4f=0,\quad\quad L_4=\left[\frac{d^2}{dr^2}r\frac{d^2}{dr^2}-\frac{d}{dr}\frac{3}{2r}\frac{d}{dr}+
2\xi-\frac{\xi^2}{r}-\frac{15}{16r^3}\right].
\label{variational-unscreened}
\end{equation}
The operator $L_4$ has the canonical form of a self-adjoint differential operator, $L_4^\dagger=L_4$,
of even order $4$ (see, for example, Chap.~4 in Ref.~\cite{Gitman}).
Since $f(r)$ describes a bound state, we search its asymptotic at large distances in the form
\begin{equation}
f(r)\sim r^\beta\exp(-s r^\alpha).
\label{ansatz-f}
\end{equation}
Then Eq.~(\ref{eq:f-critical}) for $r \to \infty$ gives an equation that determines the sought parameters $\alpha$, $\beta$, and
$s$. We find the two decreasing asymptotes with a positive real part of $s$:
\begin{equation}
\alpha=\frac{3}{4},\quad\beta=-\frac{1}{8}, \quad s=\frac{1}{3}(128\xi)^{1/4}(1\pm i).
\end{equation}
This shows that the wave function of a bound state for the bare Coulomb interaction should decrease
at infinity as $f(r) \sim e^{-s r^{3/4}}$.
The asymptotic at $r\to0$ is also easily found:
\begin{equation}
\beta=\pm1/2,\ 3/2,\ 5/2.
\label{asymptotics-zero}
\end{equation}
The trial function
\begin{equation}
f(r)=C r^{3/2}\,e^{-s r^{3/4}}
\label{variational-function-f}
\end{equation}
with variational parameter $s$ has the correct asymptotes at $r\to0$ and $r \to \infty$. The
values $\beta=\pm 1/2$ are not allowed, because the integrals in Eq.~(\ref{functional_f-4th-order}) would
be divergent at the lower limit. The integrals in Eq.~(\ref{functional_f-4th-order}) can be calculated analytically and, for $E=-\Delta$, this leads to the quadratic equation for $\xi$ whose positive root reaches the minimal value $\xi_{cr}\approx3.87$ for $s\approx1.57$. Since $\xi=(Z\alpha_g/\kappa)\sqrt{\gamma_1/\Delta}$, this result means that the critical charge of the impurity
in bilayer graphene $Z_{cr}$ for the unscreened Coulomb interaction tends to zero as $\sqrt{\Delta}$ for $\Delta \to 0$. In other words, the ground state in bilayer graphene with the zero band gap in the presence of the external Coulomb potential is always unstable with respect to the creation of electron-holes pairs. This conclusion was first obtained in Ref.~\cite{Kolomeisky} by using the semiclassical approach.

It is useful to recall that the critical charge in monolayer graphene $Z_{cr}\alpha_g \sim 1/2 + \pi^2/\log^2(\Delta)$ decreases too as the gap $\Delta \to 0$ \cite{excitonic-instability}. Crucially, it tends to a finite value $1/2$ unlike the case of bilayer graphene, where the critical charge vanishes as
the gap tends to zero. Finally, we would like to note that the wave function (\ref{variational-function-f}) is localized in terms of the dimensional variable in the region $r\le\lambda_\Delta=\hbar v_F/\sqrt{\gamma_1
\Delta}$. Since the two-band model is applicable for distances $r>r_0=\hbar v_F/\gamma_1$, this is a legitimate solution for the gap smaller than $\gamma_1$.

\subsection{Screened Coulomb potential}
\label{sec:screened}

\begin{figure}[h]
  \centering
  \includegraphics[width=8cm]{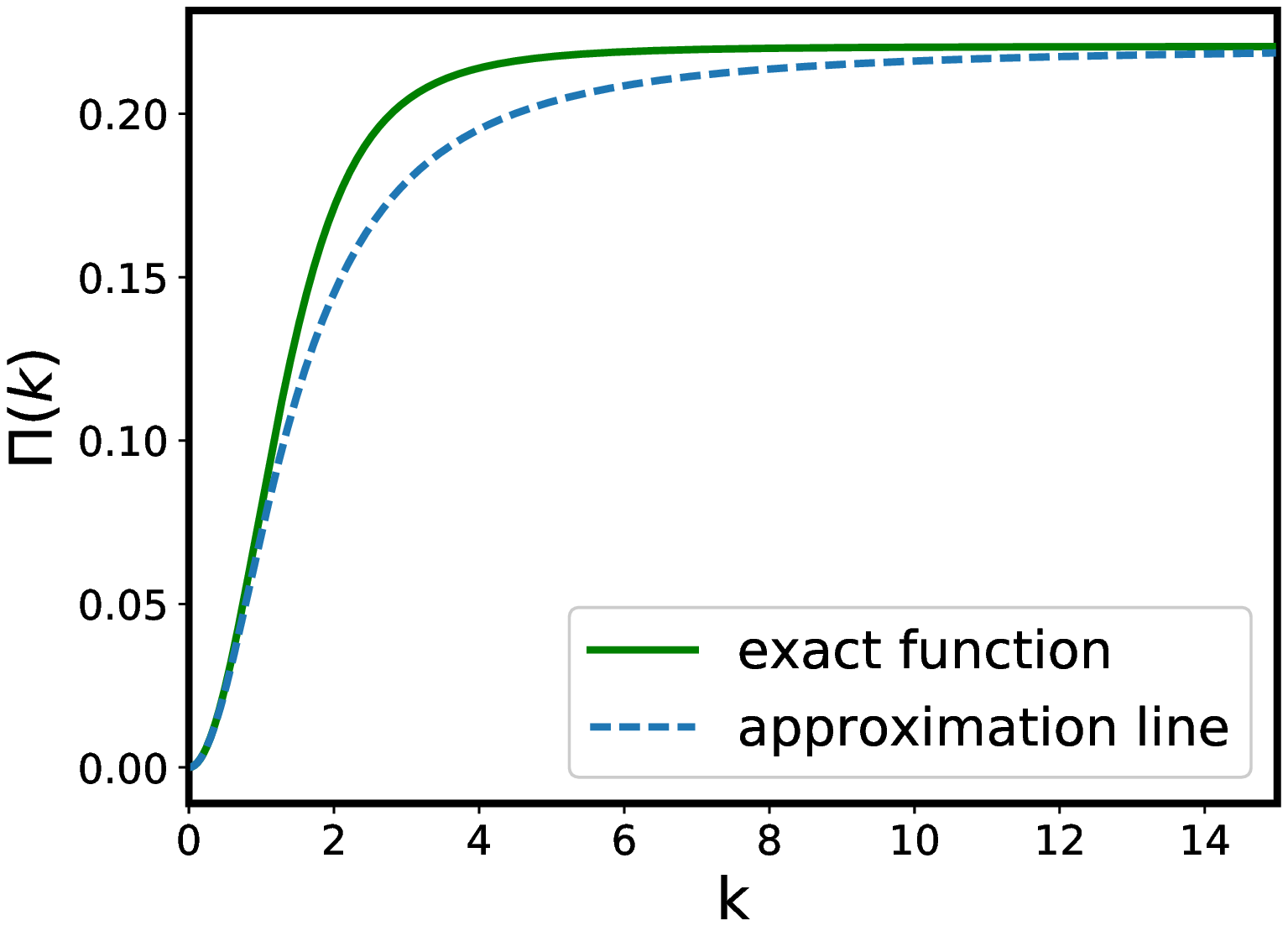}
  \includegraphics[width=7.5cm]{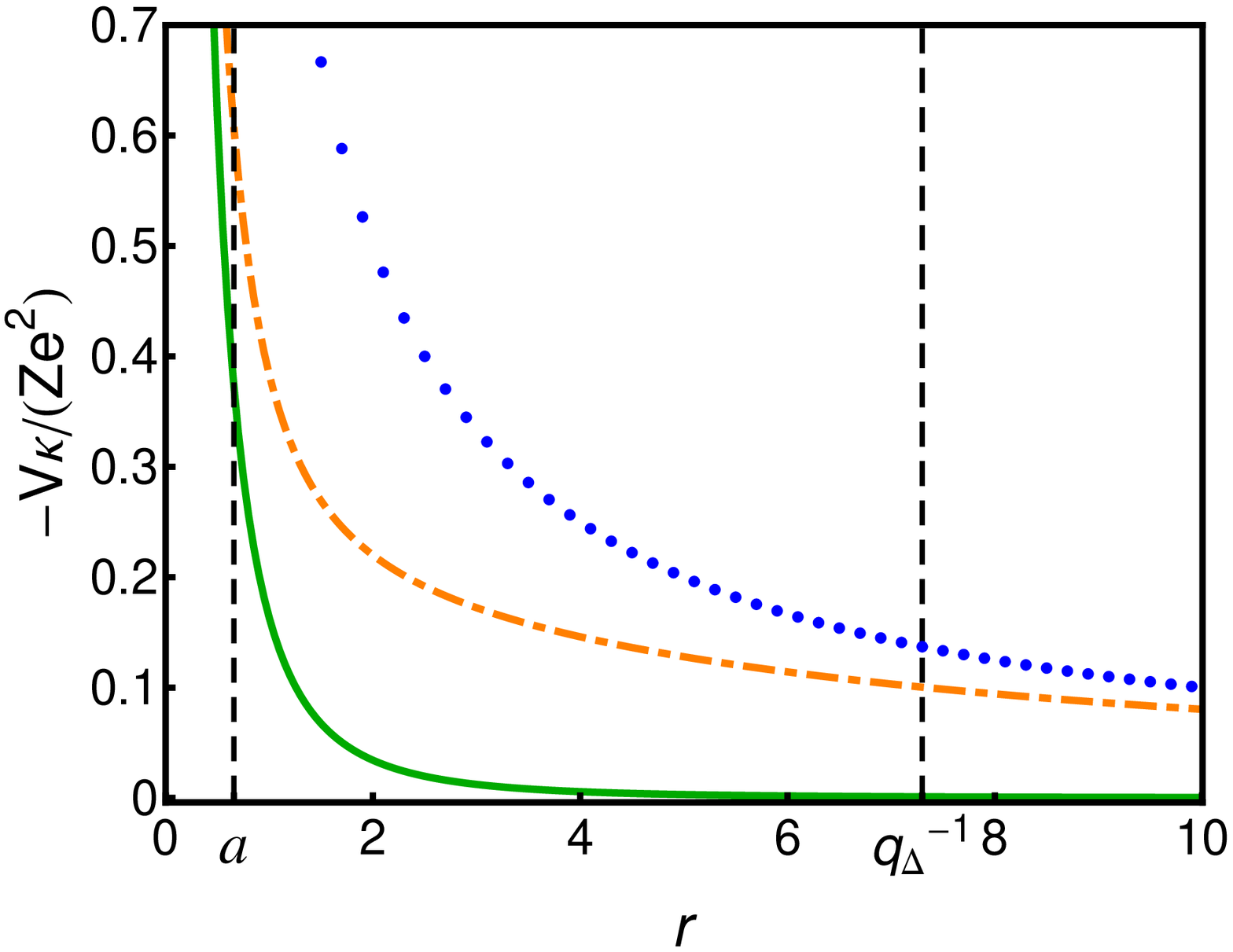}
\caption{Left panel: The dependence of the exact and approximate polarization functions on dimensionless momentum k. Right panel: The potentials $-V(r)\kappa/(Ze^2)$: the Coulomb potential $1/r$ (dotted blue line), the screened potential defined by Eq.(\ref{potential-finiteDelta}) (dash-dotted
orange line), and the screened potential in gapless bilayer graphene (\ref{potential-zeroDelta}) (solid green line) for $\alpha_g=2.19,\,\kappa=4$, and $\Delta/\gamma_1=0.1$.}
\label{potentials}
\end{figure}

Taking into account the polarization function of gapped fermions, the screened Coulomb potential $V_{eff}(r)$ in bilayer graphene equals
\begin{equation}
V_{eff}(r)=-\frac{2\pi Ze^2}{\kappa}\int\frac{d^2q}{(2\pi)^2}\frac{e^{i\mathbf{q} \mathbf{r}}}
{|\mathbf{q}|+\frac{4\pi\alpha_g}{\kappa}\Pi(q^2)},\quad q\equiv|\mathbf{q}|,
\end{equation}
where the polarization function has the form
\begin{equation}
\Pi(q^2)=\frac{2m^*}{\hbar}F\left(\frac{\hbar^2q^2}{2m^*\Delta}\right).
\end{equation}
In the one-loop approximation, $\Pi(q^2)$ is given by Eq.(\ref{exact_polariz}). Although $\Pi(q^2)$ cannot be calculated analytically, one can easily obtain its asymptotes at small and large momenta:
\begin{equation}
\Pi(q^2)\simeq\frac{\gamma_1}{\hbar v_F}\left\{\begin{array}{c}\frac{\ln2}{\pi},\quad q\to\infty, \\
\frac{(\hbar v_F)^2q^2}{3\pi\gamma_1\Delta},\quad q\to0.\end{array}\right.
\end{equation}
We will use the following approximate expression for the polarization function which nicely fits it in
the whole range of momenta:
\begin{equation}
\frac{4\pi\alpha_g}{\kappa}\Pi(q^2)=\frac{q^2}{q_\Delta+a q^2},\quad q_\Delta=\frac{3\kappa\Delta}{4\alpha_g\hbar v_F},
\quad a=\frac{\kappa\hbar v_F}{4\ln2\alpha_g\gamma_1}.
\label{polarization-fit}
\end{equation}
The comparison of the exact one-loop expression for the polarization function with the approximate expression Eq.(\ref{polarization-fit}) is presented in the left panel of Fig.\ref{potentials}. By making use of the approximate expression (\ref{polarization-fit}), we find the following analytical expression for the screened Coulomb potential:
\begin{eqnarray}
V_{eff}(r)&=&-\frac{2\pi Ze^2}{\kappa}\int\frac{d^2q}{(2\pi)^2}\frac{e^{i\mathbf{q} \mathbf{r}}}{q+q^2/(q_\Delta+aq^2)}=-\frac{Ze^2}{\kappa}\int\limits_0^\infty
\frac{d q J_0(q r)(q_\Delta+aq^2)}{q_\Delta+q+aq^2}\nonumber\\
&=&-\frac{Ze^2}{\kappa}\left[\frac{1}{r}-\frac{\pi}{2a(q_1-q_2)}
\left(q_1[\mathbf{H}_0(q_1r)-Y_0(q_1r)]-q_2[\mathbf{H}_0(q_2r)-Y_0(q_2r)]\right)\right],
\label{potential-finiteDelta}
\end{eqnarray}
where $q_{1,2}=(1\pm\sqrt{1-4a q_\Delta})/2a$ and $\mathbf{H}_0(x)$ and $Y_0(x)$ are the Struve
function and the Bessel function of the second kind, respectively (compare with similar calculations
in Ref.\cite{Pyatkovskiy}). Since $4a q_\Delta=(3/4\ln2\alpha_g^2)(\Delta/\gamma_1)\ll1$ for reasonable values of $\alpha_g$ and $\Delta$, we have $q_1\simeq 1/a$, $q_2\simeq q_\Delta$. The leading asymptotes
of this potential for small, intermediate, and large distances are given by
\begin{eqnarray}
V_{eff}(r)\simeq -\frac{Ze^2}{\kappa}\left\{\begin{array}{c}\frac{1}{r}+\frac{1}{a}\ln\frac{r e^{\gamma}}{2a},\quad
r < a,\\
\frac{1}{q^{2}_{\Delta}r^3},\quad a < r < \frac{1}{q_{\Delta}},\\
\frac{1}{r}\left(1-\frac{1}{(q_\Delta r)^2}\right),\quad r>\frac{1}{q_{\Delta}},\end{array}\right.
\label{main-domains}
\end{eqnarray}
where $\gamma\approx0.577$ is the Euler-Mascheroni constant. Equation (\ref{main-domains}) implies that the potential $V_{eff}(r)$ tends to the bare Coulomb potential for $r\to 0$ and $r\to \infty$. Such a behavior of the screened Coulomb potential is easy to understand in view of the fact that the denominator of its Fourier transform $q+4\pi\alpha_g \Pi(q^2)/\kappa = q+q^2/(q_\Delta+aq^2)$ tends to $q$ both for $q \to \infty$ and $q \to 0$. In the intermediate region $a\ll r\ll 1/q_\Delta$, the leading contribution to the screened Coulomb potential is made by the term $q_{2}[\mathbf{H}_0(q_{2}r)-Y_0(q_{2}r)]$, which falls off with distance as $1/(q_\Delta^2r^3)$. Thus,  the charge is effectively unscreened for $r<a$, it is in the region within the screening charge cloud for $a<r<1/q_{\Delta}$, and  it  is outside the screening charge cloud for $r>1/q_{\Delta}$. These regions are marked by vertical lines in the right panel of Fig.~\ref{potentials}.

In the gapless case ($q_\Delta=0$), we have
\begin{equation}
V_{eff}(r)=-\frac{Ze^2}{\kappa}\left[\frac{1}{r}-\frac{\pi}{2a}
\left(\mathbf{H}_0(r/a)-Y_0(r/a)\right)\right]\simeq -\frac{Ze^2}{\kappa}\left\{\begin{array}{c}
\frac{1}{r}+\frac{1}{a}\ln\frac{r\,e^\gamma}{2a},\quad r\ll a,\\ \frac{a^2}{r^3},
\quad r\gg a.\end{array}\right.
\label{potential-zeroDelta}
\end{equation}
Notice that in the gapless bilayer graphene the screening effects change the effective potential to
a short-range one $V_{eff}(r)\sim -1/r^3$ for large $r$. The potentials screened by gapped and gapless fermions as well as the bare Coulomb potential are plotted in the right panel of Fig.\ref{potentials}.

In order to get an analytic insight into the dependence of the critical charge on $\Delta$, it is
convenient to make the change of variable $r^{\prime}=r\sqrt{\frac{2m^*\Delta}{\hbar^2}}\equiv r/
\lambda_{\Delta}$, where $\lambda_{\Delta}=\frac{\hbar v_F}{\sqrt{\gamma_1 \Delta}}$ (this is the same change of variable as in Sec.\ref{section-model}). Then Eq.(\ref{configspace:eq}) takes the form (to simplify the notation, we switch back from $r^{\prime}$ to $r$)
\begin{equation}
\left\{
\begin{array}{c}
\left(\frac{d^2}{d\,r^2}-\frac{2\,j-1}{r}{\frac{d}{d\,r}}+\frac{j^2-1}{r^2}\right)a_j(r) = -\left(1+\epsilon
-\frac{1}{\Delta}V(\sqrt{\frac{\hbar^2 v^2_F}{\gamma_1\Delta}}\,r)\,\right)b_j(r),\\
\left({\frac{d^2}{d\,r^2}}+\frac{2\,j+1}{r}{\frac{d}{d\,r}}+{\frac{j^2-1}{r^2}}\right)b_j(r) = \left(1-\epsilon+\frac{1}{\Delta}V(\sqrt{\frac{\hbar^2 v^2_F}{\gamma_1\Delta}}\,r)\,\right)a_j(r),
\end{array}
\right.
\label{configspace:eq-rescaled}
\end{equation}
where $\epsilon=E/\Delta$.

It is instructive to consider first the Coulomb interaction in the rescaled distance. In such a case,
\begin{equation}
\frac{1}{\Delta}V_C(r\lambda_{\Delta})=-\frac{Z\alpha_g}{\kappa r}\,\sqrt{\frac{\gamma_1}{\Delta}}.
\end{equation}
Setting $\epsilon=-1$ and using $\xi=\frac{Z\alpha_g}{\kappa}\sqrt{\frac{\gamma_1}{\Delta}}$, we conclude
that $\Delta$ disappears from the corresponding differential equation. Of course, this means that
$Z_{cr} \sim \sqrt{\Delta}$ as $\Delta \to 0$.

\begin{figure}[h!]
	\centering
	\includegraphics[scale=0.5]{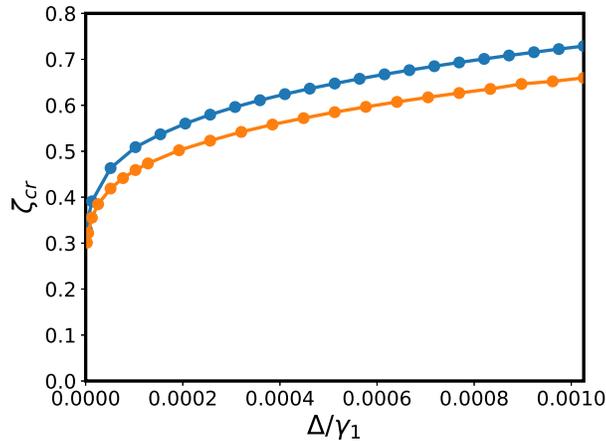}
	\caption{The critical coupling constant $\zeta_{cr}=Z_{cr}\alpha_{g}/\kappa$ as a function of $\Delta/\gamma_{1}$ obtained
	by using the variational method with the Gaussian trial function (\ref{trial_gauss})
	(lower curve) and the trial function (\ref{variational-function-f}) (upper curve). }
	\label{fig:minroots_screened}
\end{figure}

As we showed above, the screened Coulomb interaction has the following three main domains of the dependence on distance (see Eq.~(\ref{main-domains})). In terms of the rescaled distance, we find
\begin{eqnarray}
\frac{1}{\Delta}V_{eff}(r\lambda_{\Delta})\simeq -\frac{Z\alpha_g}{\kappa}\left\{\begin{array}{c}\sqrt{\frac{\gamma_1}{\Delta}}
	\frac{1}{r},\quad
	r < a/\lambda_{\Delta},\\
	\sqrt{\frac{\gamma_1}{\Delta}}\frac{1}{\lambda^2_{\Delta}}\frac{1}{q^{2}_{\Delta}r^3},\quad
	a/\lambda_{\Delta} < r < \frac{1}{\lambda_{\Delta}q_{\Delta}},\\
	\sqrt{\frac{\gamma_1}{\Delta}}\frac{1}{r}\left(1-\frac{1}{(q_\Delta\lambda_{\Delta} r)^2}\right),
\quad r>\frac{1}{\lambda_{\Delta}q_{\Delta}}.\end{array}\right.
\label{main-domains-rescaled}
\end{eqnarray}
The variational functional (\ref{functional_f-4th-order}) in terms of the rescaled distance and for $E=-\Delta$ takes the more simple form
\begin{equation}
\int\limits_0^\infty dr\frac{|L(j)f|^2}{-\frac{1}{\Delta}V_{eff}(r\lambda_{\Delta})}=\int\limits_0^\infty dr
(-2-\frac{1}{\Delta}V_{eff}(r\lambda_{\Delta}))|f(r)|^2.
\label{functional_f-4th-order-rescaled}
\end{equation}
Since $V_{eff}$ is always negative, the left-hand side of the above expression is positive. Obviously,
its right-hand side is negative if we neglect the effective potential there. This means that the variational function should be mostly localized in the region where $-\frac{1}{\Delta}V_{eff}(r\lambda_{\Delta})>2$. Since potential (\ref{main-domains-rescaled}) blows up in the first region $r < a/\lambda_{\Delta}$, it
is larger than $2$ in this region. At the right end of the second interval, i.e., at
$r=1/(\lambda_\Delta q_{\Delta})$, the rescaled effective potential takes a finite value which
is, in general, larger than 2. This means that the first and second regions are important for
the supercritical instability and should be definitely taken into account. On the other hand, potential (\ref{main-domains-rescaled}) decreases as $1/r$ in the third interval and $-\frac{1}{\Delta}V_{eff}(r\lambda_{\Delta})$ becomes less than 2 in this region. This
suggests that the best choice for the variational function $f(r)$ would be a function which rapidly decreases for $r > \frac{1}{\lambda_{\Delta}q_{\Delta}}$. For example, the simplest choice would be a Gaussian, which is mostly localized in the second distance domain
\begin{equation}
\label{trial_gauss}
f_G(r)=Cr^{3/2}e^{-br^{2}/(\lambda_{\Delta}q_{\Delta})^2},
\end{equation}
where $b$ is a dimensionless variational parameter. The prefactor $r^{3/2}$ is added in order to obtain
the correct asymptotic behavior of the trial function for $r \to 0$, which follows from Eq.~(\ref{variational-function-f}). We use also the trial function (\ref{variational-function-f}), which
like the Gaussian ansatz rapidly decreases with $r$. The integrals in the variational functional (\ref{functional_f-4th-order-rescaled}) with the screened Coulomb potential cannot be  evaluated analytically; therefore, we calculate them numerically. The corresponding results for the critical coupling constant as a function of $\Delta/\gamma_1$ for the trial functions (\ref{variational-function-f}) and (\ref{trial_gauss}) are shown in Fig.\ref{fig:minroots_screened}. As one can see from this figure, the trial functions  give quite similar results and $\zeta_{cr}=Z_{cr}\alpha_{g}/\kappa$ in both cases tends to a finite value for $\Delta=0$. For example, for $\Delta=0.001$ meV, $\zeta_{cr}=0.33$ for the trial function (\ref{variational-function-f}) and $\zeta_{cr}=0.30$ for the Gaussian trial function (\ref{trial_gauss}). Note that the Gaussian trial function (\ref{trial_gauss}) fares better because it gives a smaller value of the critical coupling constant for all considered values of $\Delta$.

\section{Numerical results}
\label{section-numerical}

In order to find a numerical solution to the system of equations (\ref{system}), we split the momentum interval $(0,\Lambda)$ into $N$ equal intervals and approximate the integrals by the sums of values of integrands at the ends of intervals multiplied by the weight function $w_i$ of the Newton-Cotes formula
of the fifth order:
\begin{equation}
{\int\limits_{0}^{\Lambda}dq q [K_{j}(k,q)-\delta V_{j}(k,q)]b_j(q)}=\sum_{i=0}^{N}q_{i} w_{i} [K_{j}(k,q_{i})-\delta V_{j}(k,q_{i})]b_j(q_{i}).
\label{kernelsum}
\end{equation}
The kernel $K_{j}(k,q)$ defined in Eq.~(\ref{kernel}) has a logarithmic singularity at $q=k$. In order to deal with this singularity, we use the following regularization for the first term in the square brackets
on the left-hand side of Eq.(\ref{kernelsum}):
   	\begin{equation}
   	\label{reg}
   	{\int\limits_{0}^{\Lambda}dq q K_{j}(k,q)b_j(q)}=\sum_{i=0}^{N}q_{i} w_{i} K_{j}(k,q_{i})[b_j(q_{i})-b_j(k)]
   	+b_j(k){\int\limits_{0}^{\Lambda}dq q K_{j}(k,q)}.
   	\end{equation}
The last integral in Eq.(\ref{reg}) is not singular and could be expressed through the elliptic integrals. In the Appendix we present the integrals for $j=0$, 1, 2, and 3 (see Eqs.~(\ref{int0})-(\ref{int3}), respectively). Using these expressions we can calculate the critical charge for the total angular momentum from $j=-2$ to $j=2$. For monolayer graphene, we checked that our numerical algorithm gives the same spectrum as the shooting method. The quadrature step $h$ corresponds to the inverse of the size of a graphene disk; therefore, the energy spectrum of the system is discrete and the upper and lower continua existing in an infinite system appear now as the sets of closely situated discrete levels with the distance between them proportional to $h$.  The energy levels of the corresponding upper and lower quasicontinua drift very slowly as the impurity charge increases. On the other hand, the bound levels inside the band gap shift towards the lower quasicontinuum much faster. Therefore, there exists a critical value of the impurity charge when the lowest-energy bound state approaches the highest-energy state of the lower quasicontinuum. We define this value as the critical charge. For different values of total angular momentum $j$, we find the different values of the critical charge (see the left panel in Fig.~\ref{FH_bilayer_spectrum_m-2}). The minimal value of the critical charge is obtained for $j=1$. This finding agrees with the results in Ref.~\cite{Shytov-states}. Therefore, all numerical computations in this section will be performed for this value of the total momentum.

\begin{figure}[h!]
	\centering
	\includegraphics[scale = 0.5]{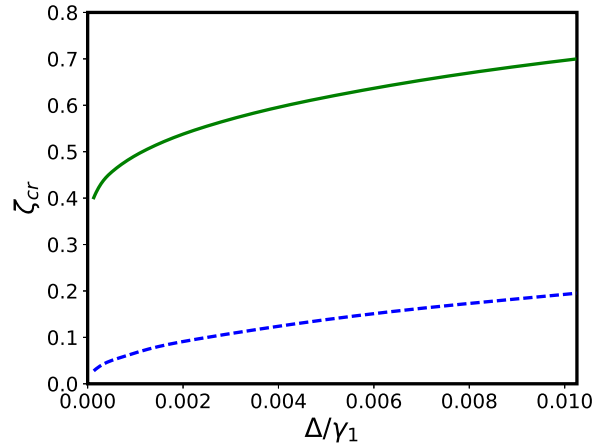}
	\caption{The critical coupling constant when the lowest energy bound state enters the lower continuum as a function of dimensionless gap $\Delta/\gamma_1$ with (green solid line) and without (blue dashed line) the screening effects taken into account. The corresponding energy levels of the electron bound states are obtained from the numerical solution to Eq.~(\ref{system}). }
    \label{spectrum_fig}
\end{figure}
	
The dependence of the critical coupling constant on gap $\Delta$ is plotted in Fig.~\ref{spectrum_fig}
in the cases where the screening effects are absent and taken into account. Clearly, the critical charge decreases in both cases as the gap decreases. According to our numerical calculations, the critical charge in the absence of the screening effects is very well approximated by the function
$\zeta^{fit}_{cr}=1.88\sqrt{\Delta/\gamma_{1}}$. Such a gap dependence excellently agrees with
the conclusion made in Sec.~\ref{section-variational}. If the screening effects are taken into account,
then the critical coupling constant does not tend to zero as $\Delta$ goes to zero. The results of the numerical solution to system (\ref{system}) are quite similar to those obtained in Sec.~\ref{sec:screened}
by using the variational method (compare with Fig.~\ref{fig:minroots_screened}) and can be fitted by the
function
\begin{equation}
	\zeta^{fit}_{cr}=0.36+4.32\sqrt{\Delta/\gamma_{1}}.
\label{eq:zeta_screened_fitting}	
\end{equation}

In order to quantify the localization properties of the wave function of the electron bound state in
the near-critical regime and check the consistency of the use of the two-band model for the study of the supercritical instability in bilayer graphene, we plot for $Z \approx Z_{cr}$ in the right panel of Fig.~\ref{FH_bilayer_spectrum_m-2} the square of the wave function for $j=1$ in the two-band model in momentum representation $W(k)=2\pi k[a_{j}^2(k)+b_{j}^2(k)]/N$ multiplied by the weight factor $2\pi k$, where $N$ is the normalization constant $N=\int_{0}^{\Lambda}2\pi k [a_j^2(k)+b_j^2(k)]\,dk$. Obviously,
if the wave function is localized in momentum space in the region of momenta $k \ll \Lambda$, then the use
of the low-energy model for the description of the supercritical phenomena is consistent. According to
the right panel in Fig.~\ref{FH_bilayer_spectrum_m-2}, the maximum of the wave function corresponds to
$k \approx 0.1$. This value is 10 times less than the cutoff of the two-band model. This result suggests that the low-energy two-band model consistently describes the supercritical behavior in bilayer graphene.

\begin{figure}[h!]
	\centering
\includegraphics[scale = 0.48]{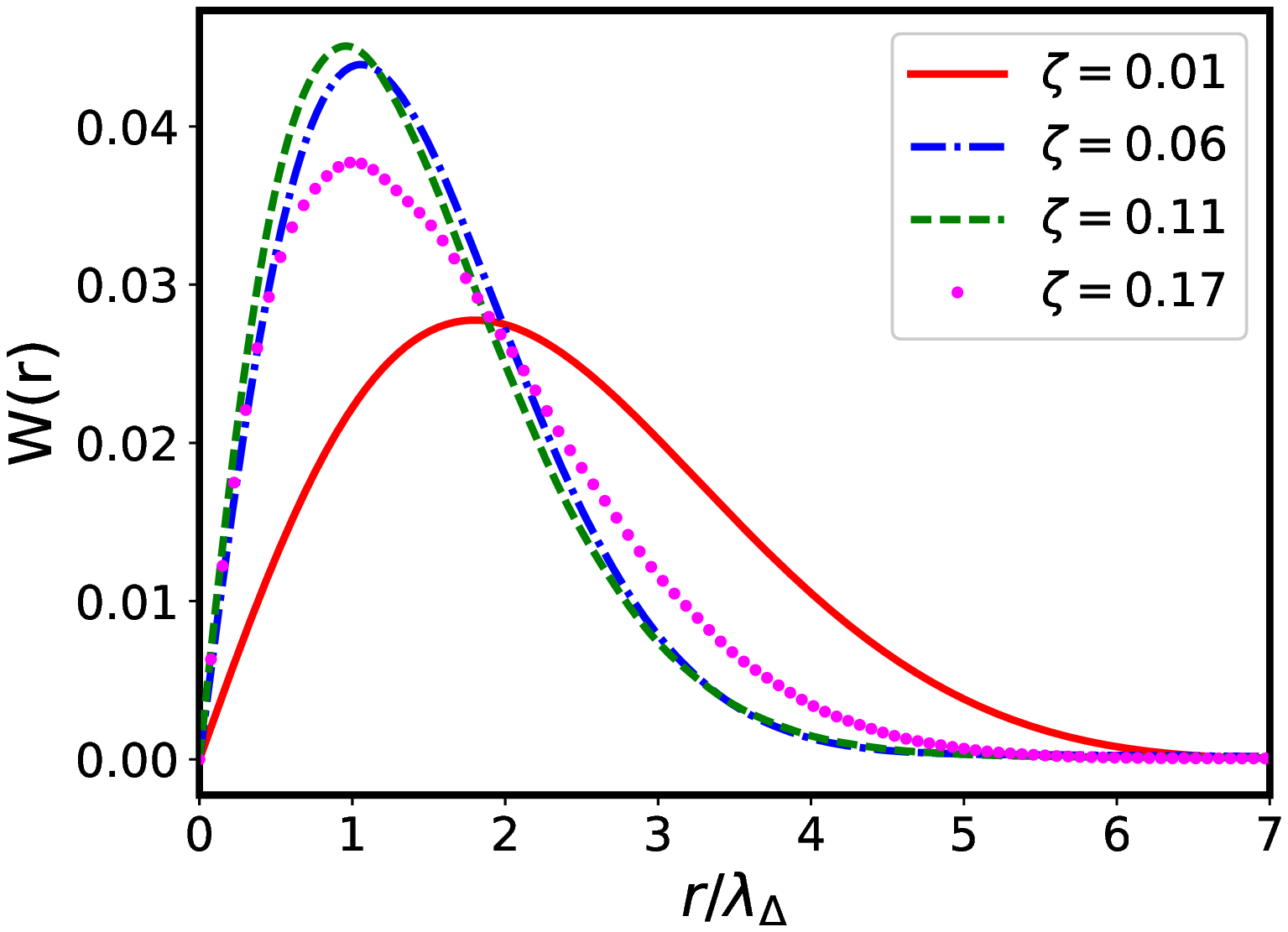}
\includegraphics[scale = 0.4]{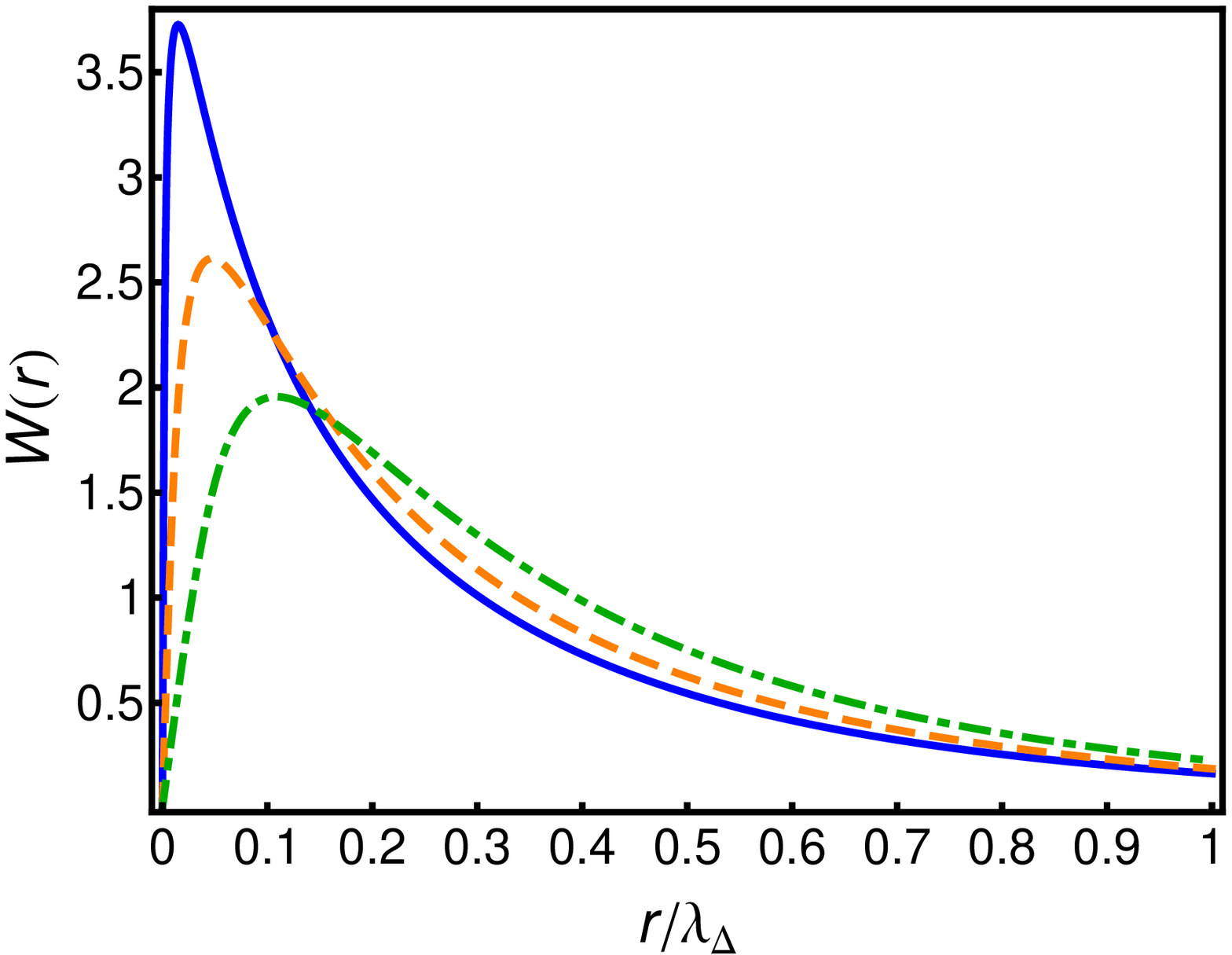}
\caption{Left panel: The electron density of the lowest energy state in bilayer graphene as a function of distance for several values of the coupling constant. Right panel: Radial distribution of the electron density  in monolayer graphene for near-critical charges of the impurity for three different values of regularization parameter: $R=0.001 \lambda_{\Delta}$ (blue solid line), $R=0.01 \lambda_{\Delta}$ (orange dashed line), and $R=0.05 \lambda_{\Delta}$ (green dash-dotted line).
}
\label{comparison}
\end{figure}

As to the phenomenon of the fall to center, it is obviously absent in the two-band model where the more singular scaling of the kinetic energy with distance prevents the electron from falling to the center and eliminates the need to regularize the Coulomb potential. To demonstrate it explicitly, we plot in the left panel of Fig.~\ref{comparison} the square of the wave function of the electron bound state of the lowest energy in bilayer graphene as a function of distance for several values of the coupling constant. The components $a_j(r),\ b_j(r)$ of a wave function (\ref{substitution}) in configuration space can be evaluated from the components $a_j(k),\ b_j(k)$ (\ref{spinor}) in momentum space  by means of the Hankel transform,
\begin{equation}
a_j(r)=i^{j-1}\int\limits_0^\infty \frac{dk}{2\pi} k J_{j-1}(k r)a_{j}(k),\quad b_j(r)=i^{j+1}\int\limits_0^\infty \frac{dk}{2\pi} k J_{j+1}(k r)b_{j}(k),
\end{equation}
where $J_\nu(z)$ is the Bessel function. It is seen that the wave function is still at a significant distance from $r=0$ when the exciton instability $\zeta_{cr}$ is achieved. For comparison, we plot in the right panel of the same figure the square of the wave function in monolayer graphene for near-critical charges of the impurity $\zeta_{cr}-\zeta<0.01$ for different values of the regularization parameter $R$. When $R$ decreases the electron wave function shrinks towards the impurity (fall to center).

Although our results show that the analysis of the supercritical instability is consistent in the two-band model, it is still necessary to study the supercritical instability in the four-band model of bilayer graphene. The point is that the quadratic energy dispersion in the two-band model is replaced by the linear energy dispersion as in monolayer graphene for momenta larger than $\gamma_1/(4v_F)$. Therefore, it
is possible, in principle, that the supercritical instability will be affected by the electron dynamics at short distances. The four-band model whose energy dispersion smoothly interpolates between the low-energy quadratic and high-energy linear-in-momentum energy dispersion allows us to investigate whether the conclusions made in the present section survive in the four-band model.

\section{Four-band model}
\label{four-band-model}

\begin{figure}[h!]
	\centering
\includegraphics[scale = 0.5]{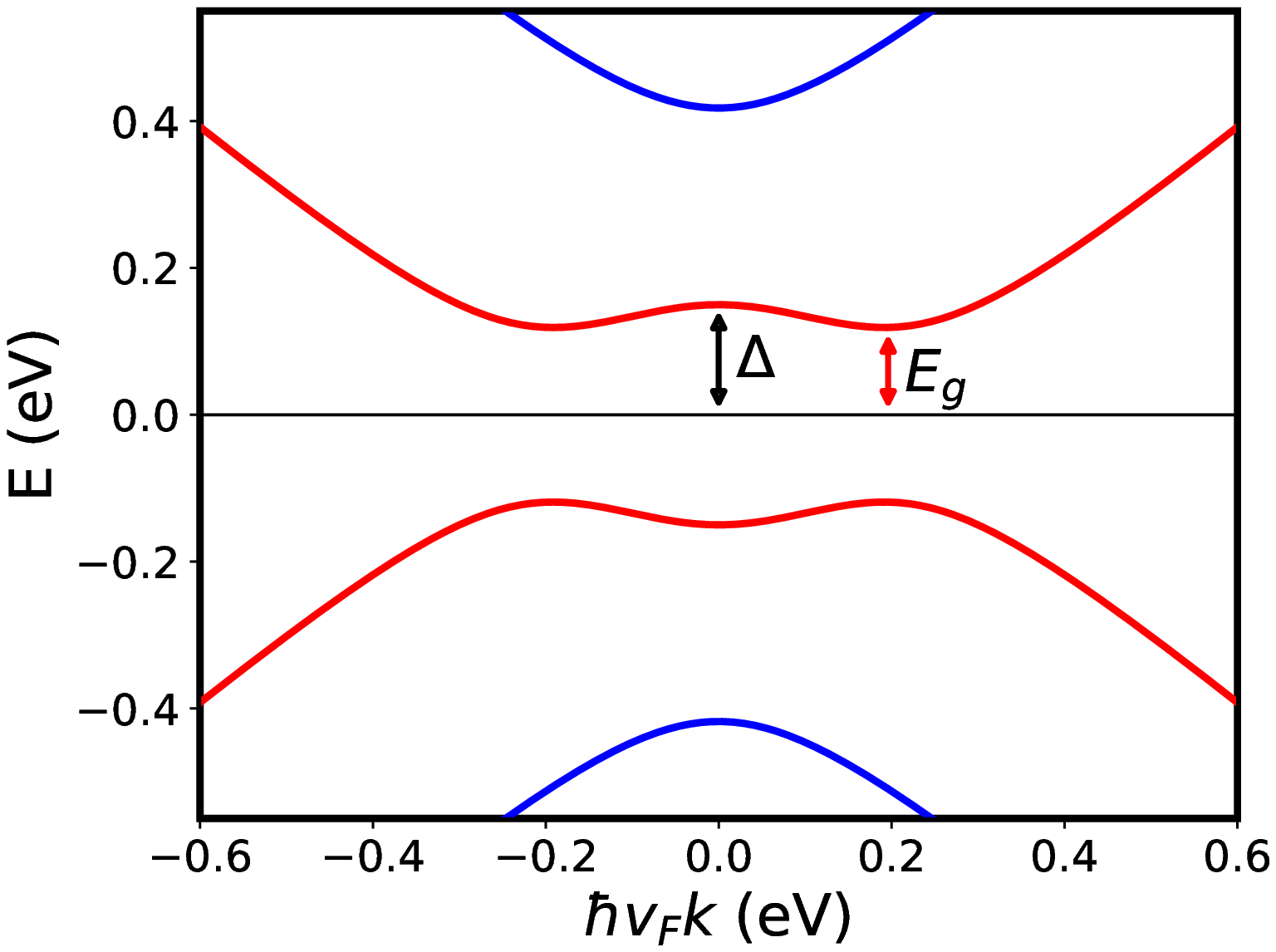}
\includegraphics[scale = 0.5]{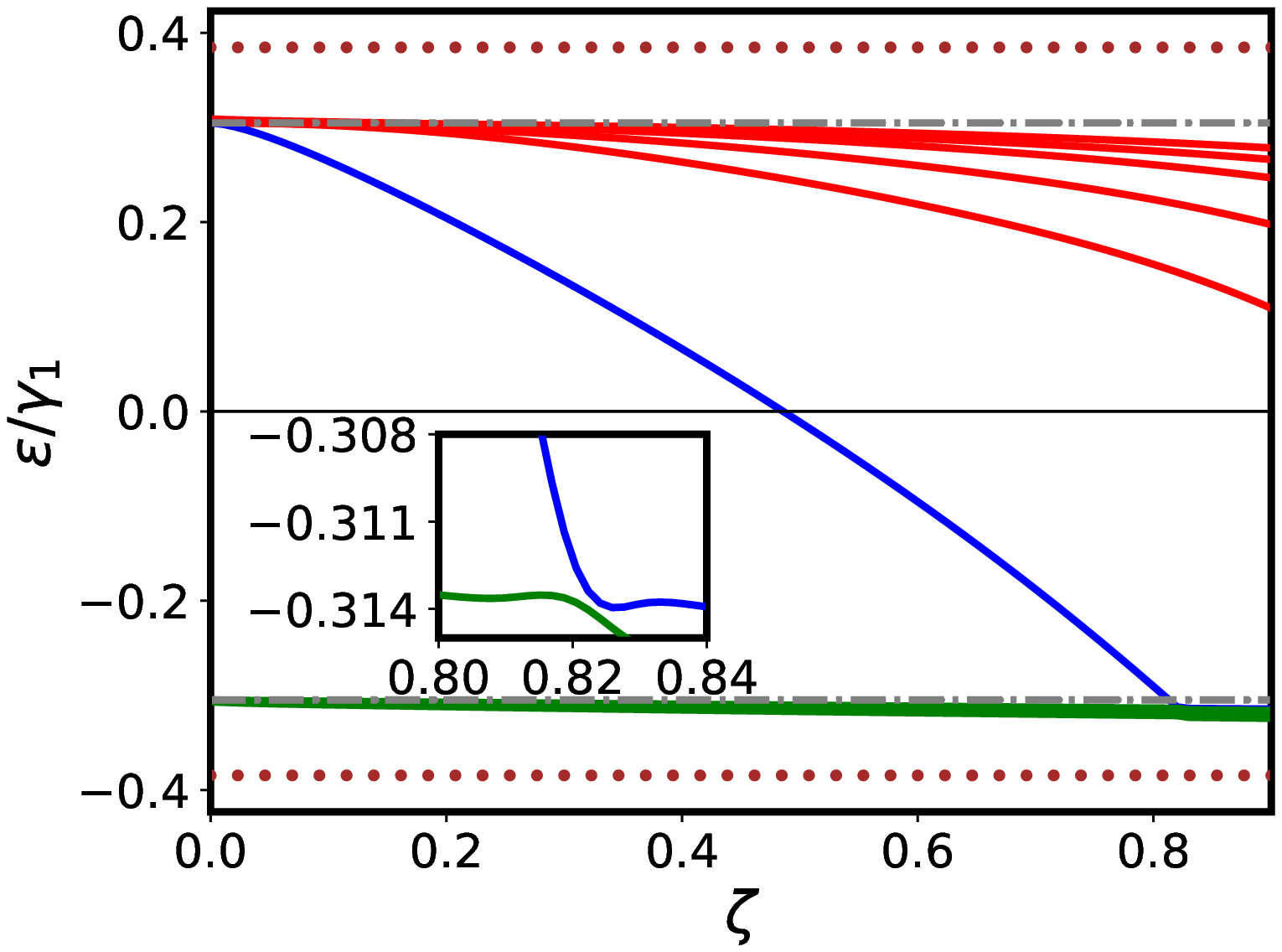}
\caption{Left panel: The energy spectrum of gapped bilayer graphene for $\Delta=150\, {\rm meV}$. Right panel: The midgap electron bound states in the field of a charge impurity in bilayer graphene. The momentum cut-off $\Lambda_4=10$ and energy gap $\Delta=150$\,meV are used in the four-band model. The electron bound states descend from $E_{g}$ which equals $118.9$ meV in the case under consideration. Inset of the right panel:
A zoom-in of the region where the bound state of the lowest energy enters the states of the lower quasicontinuum.}
\label{spectrum-bilayer-graphene}
\end{figure}

By making use of the tight-binding model of graphite and the Slonczewski--Weiss--McClure parametrization \cite{Dresselhaus}, the four-band model which describes quasiparticles in the Bernal-stacked bilayer graphene was found in Ref.\cite{bilayer}. The Hamiltonian of this model operates in the space of 4-component wave functions $\Psi^{T}_{K}=(\Psi_{A_1},\Psi_{B_2},\Psi_{A_2},\Psi_{B_1})$ in the valley $K$ and
$\Psi^{T}_{K'}=(\Psi_{B_2},\Psi_{A_1},\Psi_{B_1},\Psi_{A_2})$ in the valley $K'$ and reads
\begin{equation}
	H^{(0)} = \eta\left(\begin{array}{cccc}
	\Delta & 0 & 0 & v_{\rm F} p_{-}\\
	0 & -\Delta & v_{\rm F} p_{+} & 0 \\
	0 & v_{\rm F} p_{-} & -\Delta & \eta \gamma_1\\
	v_{\rm F} p_{+} & 0 & \eta \gamma_1 & \Delta\\
	\end{array}
	\right),
	\label{free-hamiltonian}
\end{equation}
where $p_{\pm}=p_{x}\pm i p_{y}$, $\mathbf{p}$ is the momentum measured with respect to the corresponding Dirac point, $\Delta$ is the quasiparticle gap, and $\eta=+1$ ($\eta=-1$) labels the valley $K$ ($K'$).
It is straightforward to show that this Hamiltonian defines the following four energy bands:
\begin{equation}
E^2(\mathbf{p})=\frac{\gamma^2_1}{2}+\Delta^2+v^2_F\mathbf{p}^2\,\pm\,\sqrt{\frac{\gamma^4_1}{4}
+v^2_F\mathbf{p}^2(\gamma^2_1+4\Delta^2)}.
\label{energy-found-band-model}
\end{equation}
The positive sign describes two high-energy bands whose energy square, obviously, is always larger than $\gamma^2_1$. The negative sign corresponds to the low-energy bands whose energy dispersion smoothly interpolates between that of the two-band model considered in the previous section and the energy relation of monolayer graphene at energies larger than $\gamma_1/2$. The spectrum of quasiparticle excitations in gapped bilayer graphene is shown in the left panel of Fig.\ref{spectrum-bilayer-graphene}.
It is interesting that if the gap is of order $\gamma_1$, then the energy dispersion relation of the
low-energy band resembles the Mexican hat \cite{Carbotte_Nikol} (for $\Delta \ll \gamma_1$, the energy dispersion relation is indistinguishable from the quadratic one). The energy gap between the two low-energy bands in the four-band model equals $2E_{g}$, where
$E_{g}=\frac{\Delta\gamma_{1}}{\sqrt{4\Delta^2+\gamma^{2}_{1}}}$ (obviously, $E_{g} \approx \Delta$ for $\Delta \ll \gamma_1$).

Since the free Hamiltonian (\ref{free-hamiltonian}) and the Coulomb interaction are rotation invariant,
the angle and the absolute value of momentum variables can be separated in the eigenvalue problem for the electrons in the field of a charged impurity in bilayer graphene. It is convenient in this section to express energy in terms of $\gamma_1$, distances in terms of $\lambda_{\gamma_{1}}=v_{\rm F}\hbar/\gamma_1$, and momenta in terms of $\lambda^{-1}_{\gamma_{1}}$. Then the eigenvalue problem for the electrons in the Coulomb field of a charged impurity in bilayer graphene takes the following form in polar coordinates $(k,\theta)$:
\begin{equation}
	\left(\begin{array}{cccc}
	\Delta/\gamma_1 & 0 & 0 & k e^{-i\theta}\\
	0 & -\Delta/\gamma_1 & k e^{i\theta} & 0 \\
	0 & k e^{-i\theta} & -\Delta/\gamma_1 &  1\\
	k e^{i\theta} & 0 &  1 & \Delta/\gamma_1\\
	\end{array}
	\right)\Psi(\mathbf{k}) + \int\!\frac{d^2q}{(2\pi)^{2}}\,V(\mathbf{k}-\mathbf{q})\Psi(\mathbf{q}) = \epsilon\Psi(\mathbf{k}),
\end{equation}
where $\epsilon=E/\gamma_1$ and $V(\mathbf{k})$ is the bare Coulomb interaction. We established in the previous section that the supercritical instability is self-consistent in the two-band model and the fall-to-center is not realized. However, it is necessary to check that the supercritical instability is
not affected by the electron dynamics at short distances where the electron dispersion relation is linear. In this section, we perform the analysis of the supercritical charge instability in the four-band model.
For simplicity, we consider the bare Coulomb interaction. Since we will see that the analysis in the four-band models reproduces very well the results obtained in the two-band model, there is no need to analyze a more complicated case of the screened Coulomb interaction. Choosing the spinor wave function in the form
\begin{equation}
\Psi(\mathbf{k})=\left(
\begin{array}{c}
a_j(k) e^{i(j-1)\theta}\\
b_j(k) e^{i(j+1)\theta}\\
c_j(k) e^{ij\theta}\\
d_j(k) e^{ij\theta}
\end{array}
\right),
\end{equation}
we obtain the following system of equations for the coefficient functions, which depend only on the
modulus of momentum:
\begin{equation}
	\begin{array}{c}
	(\Delta/\gamma_1)a_j(k)+kd_j(k)-\frac{\zeta}{2\pi}\int\limits_{0}^{\Lambda_4}dq q
K_{j-1}(k, q) a_j(q) = \epsilon a_j(k),\\
	-(\Delta/\gamma_1)b_j(k) + kc_j(k) - \frac{\zeta}{2\pi}\int\limits_{0}^{\Lambda_4}dq q
K_{j+1}(k, q) b_j(q) = \epsilon b_j(k),\\
	kb_j(k) - (\Delta/\gamma_1)c_j(k) +d_j(k) - \frac{\zeta}{2\pi}\int\limits_{0}^{\Lambda_4}dq q K_{j}(k, q) c_j(q) = \epsilon c_j(k),\\
	ka_j(k)+c_j(k)+(\Delta/\gamma_1)d_j(k) - \frac{\zeta}{2\pi}\int\limits_{0}^{\Lambda_4}dq q K_{j}(k, q) d_j(q) = \epsilon d_j(k),
	\end{array}
\label{system-numerical}
\end{equation}
where $\zeta=Z\alpha_{g}/\kappa$ and the dimensionless cutoff $\Lambda_4 \equiv \lambda_{\gamma_{1}}
\pi/a\approx 10$ of the four-band model is twenty times larger than the cutoff $\Lambda$ of the two-band model which equals $1/2$ in terms of $\lambda^{-1}_{\gamma_1}$. This cutoff describes the whole Brillouin zone in graphene. Numerically solving the system of equations (\ref{system-numerical}), we plot the spectrum of the electron bound states in the field of a charged impurity in the four-band model of bilayer graphene in the right panel of Fig.~\ref{spectrum-bilayer-graphene}. According to this panel, the electron bound states descend from $E_{g}$ (plotted as the horizontal dash-dotted gray line) rather than $\Delta$ (plotted as the horizontal dashed brown line). This is in agreement with the results of Ref.\cite{Skinner} where bound states of the electron in the field of a Coulomb impurity were studied near the bottom of the Mexican hat structure in the dispersion relation. The charge of impurity becomes supercritical when the lowest bound state approaches $-E_{g}$.

We would like to emphasize one more feature connected with the right panel in Fig.~\ref{spectrum-bilayer-graphene}. When the bound state approaches the lower continuum, according to
the inset of this panel, it does not cross its levels. Instead the level repulsion is realized in agreement with the Wigner--von Neumann theorem and the results in Ref.\cite{Greiner} in the case of a discrete spectrum. This means that the bound state hybridizes with the hole continuum and, as we mentioned in the Introduction, becomes a resonance causing a supercritical instability.

\begin{figure}[h!]
	\centering
	\includegraphics[scale = 0.52]{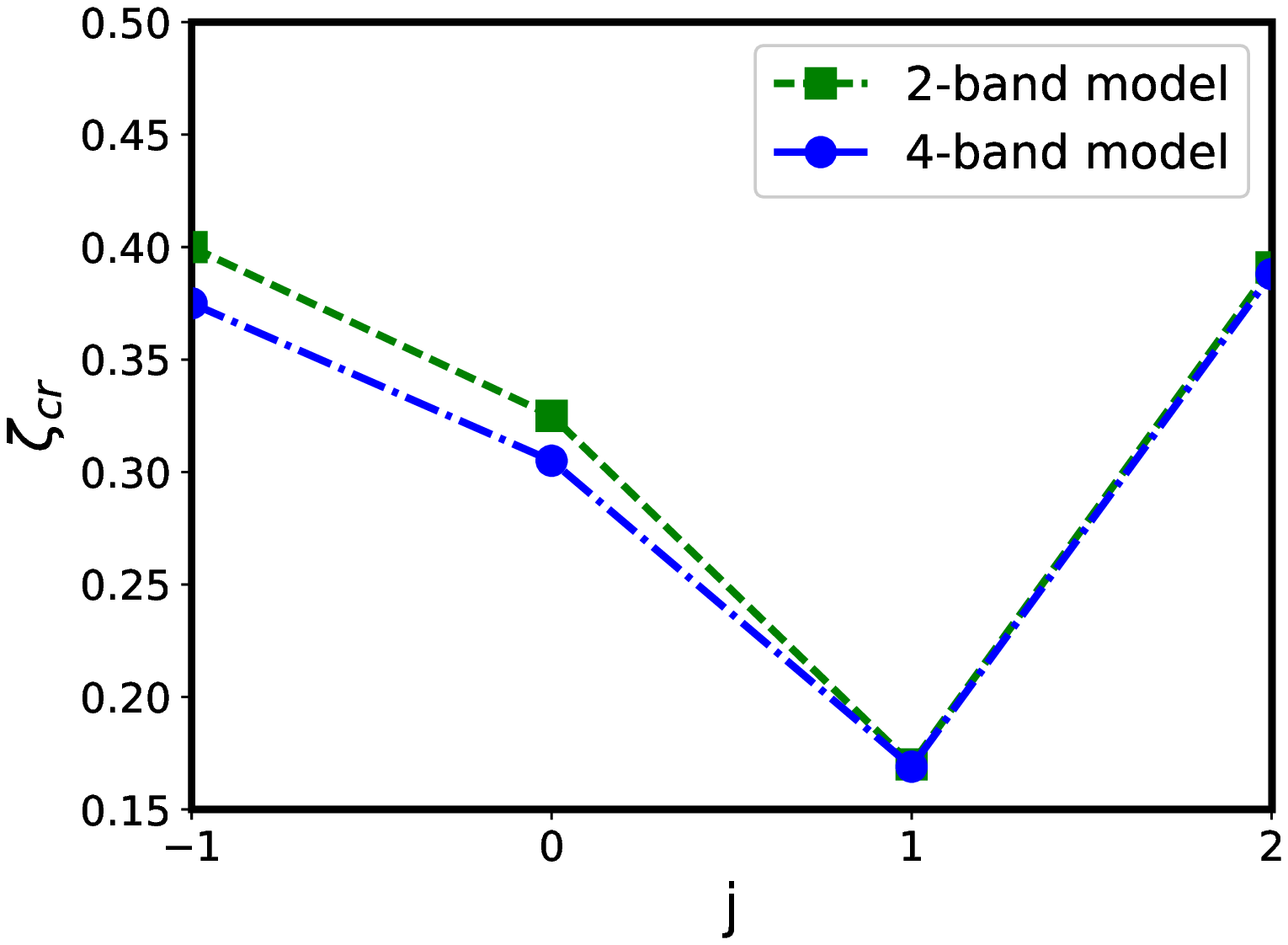}
\includegraphics[scale = 0.52]{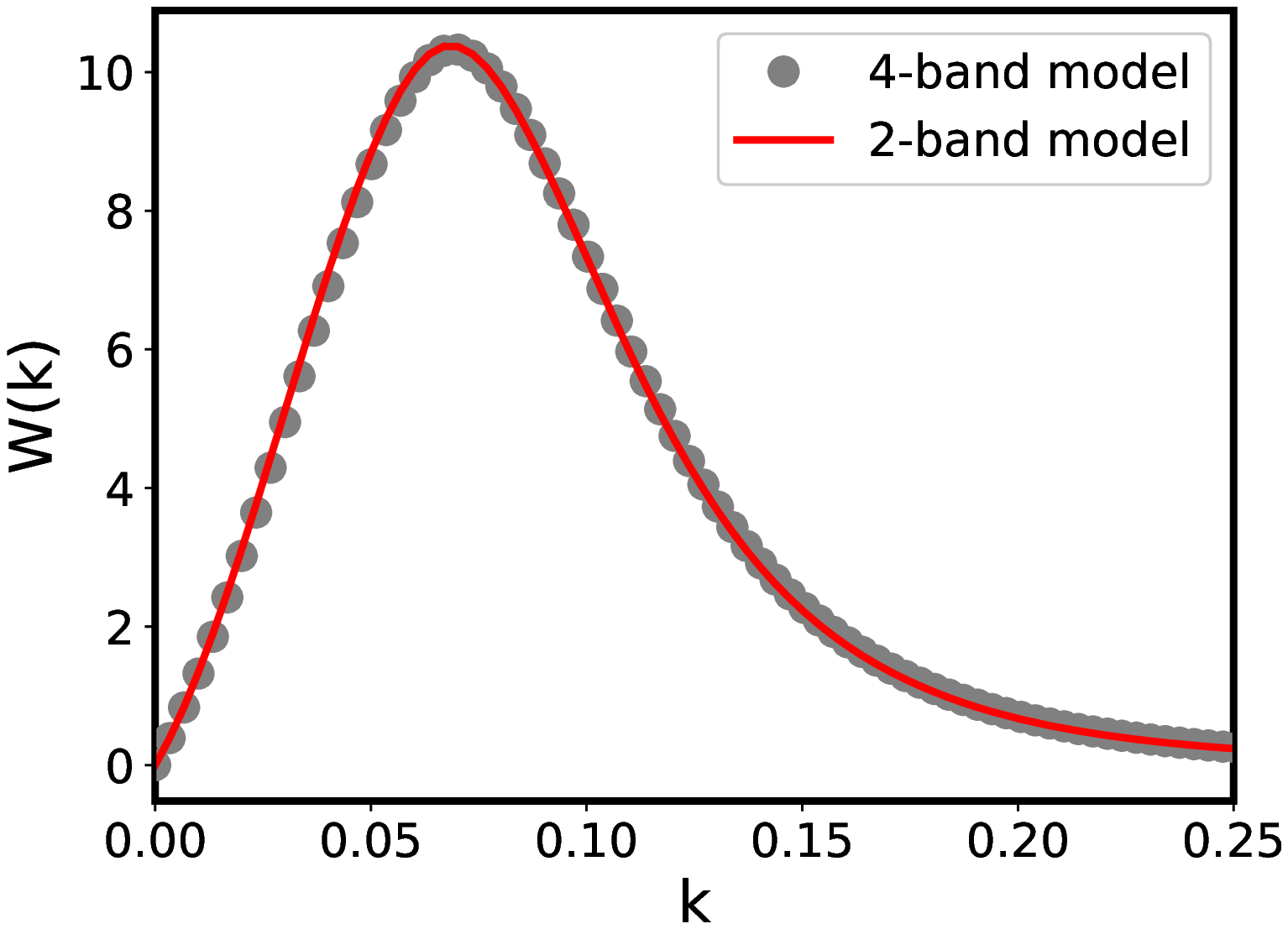}
	\caption{Left panel: The critical coupling constant as a function of total angular momentum $j$ for $\Delta=3$ meV. Right panel: The square of the wave function of the electron bound state for the near-critical charge $\zeta=\frac{Z\alpha_g}{\kappa}=0.17$ and the gap $\Delta=3$ meV plotted for the two-band and four-band models as functions of dimensionless momentum, which is expressed in terms of $1/\lambda_{\gamma_{1}}$. }
	\label{FH_bilayer_spectrum_m-2}
\end{figure}

It is instructive to compare the results of calculations performed in the two-band and four-band models. According to the left panel in Fig.~\ref{FH_bilayer_spectrum_m-2}, the critical charge is rather close
for different values of total momentum $j$ in both models. Note that the higher band in the four-band model practically does not influence the values of the critical charges for momenta $j=1,\,2$ while it gives slightly lower critical charges for $j=-1,\,0$. This is related to the fact that the wave functions with $j=-1,\,0$ in the two-band model are localized at larger momenta than those with $j=1,\,2$, and larger momenta are more influenced by the higher band in four-band model. Thus, we can conclude that for small values of
the energy gap the modification of the model does not make any noticeable changes in the critical
values. In order to check the localization properties of wave functions in the near-critical regime, we
plot in the right panel of the same figure the square of the wave function of the electron bound state with $j=1$ in momentum space $W(k)=2\pi k[a_j^2(k)+b_j^2(k)]/N$ in the two- and four-band models. Clearly, the wave functions are localized in the same domain of momentum space in these models. Thus, the results
of this section prove that the low-energy two-band model gives fairly accurate results for the Coulomb center problem in the case of small gaps.

\section{Summary}
\label{section-conclusion}

In this paper, we studied the supercritical charge instability for the electrons in the field of a charged
impurity in gapped bilayer graphene. We showed that the lowest-energy electron bound state descends
from the upper continuum and dives into the lower continuum as the charge of the impurity increases.
We found that the critical charge differs from that in monolayer graphene and depends on the gap. The screening effects play a profound role in the study of the supercritical instability in bilayer graphene.
If they are neglected, then the critical coupling constant tends to zero as the square root of the gap
when the latter vanishes. This result agrees with the conclusion obtained by using the semiclassical approach in Ref.~\cite{Kolomeisky} that the ground state of the Coulomb center problem in gapless bilayer graphene is always unstable with respect to the creation of electron-hole pairs. If the Coulomb interaction is screened, then the critical coupling tends to a finite value as the gap goes to zero.

Since the wave function of the electron bound state does not shrink to the charged impurity as the charge
of the latter increases, the fall to center does not occur. This means that the supercritical charge instability in bilayer graphene is not connected with the phenomenon of the fall to center. The underlying physical reason for this is that the electron kinetic energy and the Coulomb interaction scale differently in bilayer graphene unlike the case of monolayer graphene where both quantities scale in the same way.

We showed that the wave function is localized in momentum space in the region with momenta ten times smaller than the cutoff of the two-band model. This means that the supercritical instability in bilayer graphene takes place in the low-energy regime, where the band spectrum is quadratic. Our analysis
in Sec.~III shows also that the gap plays a very important role in the screening of the Coulomb
potential and in the localization properties of the wave function of the lowest-energy bound state.

\begin{acknowledgments}
The work of E.V.G. and V.P.G. is supported by the Program of Fundamental Research of the Physics and Astronomy Division of the NAS of Ukraine. V.P.G. acknowledges the support of the RISE Project CoExAN GA644076.
\end{acknowledgments}
\vspace{5mm}

\appendix
\section{}
\label{A}

In this appendix we collect the expressions for the kernels $K_{j}(k,q)$ defined in Eq.~(\ref{kernel}) in terms of complete elliptic integrals. We first write
\begin{equation}
K_j(k,q)=\frac{2}{\sqrt{k q}}Q_{|j|-1/2}\left(\frac{k^2+q^2}{2k q}\right)=\frac{2}{\sqrt{k q}}
Q_{|j|-1/2}\left(\cosh\eta\right),\quad \eta=\ln\frac{k}{q}.
\end{equation}
For the Legendre function $Q_\lambda(z)$ we have the recurrence relations (see Eq.(8.832.4) in Ref.\cite{Gradstein})
\begin{equation}
Q_\lambda(z)=\frac{2\lambda-1}{\lambda}z Q_{\lambda-1}(z)-\frac{\lambda-1}{\lambda}Q_{\lambda-2}(z).
\label{recursion}
\end{equation}
Starting from the particular values
\begin{equation}
Q_{-1/2}(\cosh\eta)=2e^{-|\eta|/2}{\rm K}\left(e^{-|\eta|}\right),\quad Q_{1/2}(\cosh\eta)=2e^{|\eta|/2}\left[{\rm K}\left(e^{-|\eta|}\right)-{\rm E}\left(e^{-|\eta|}\right)\right],
\end{equation}
where ${\rm K}(x)$ and ${\rm E}(x)$ are the complete elliptic integrals of the first and second kind, respectively, all other values of $Q_{|j|-1/2}(\cosh\eta)$ can be obtained using recursions (\ref{recursion}). Hence all kernels $K_j(k,q)$ are expressed in terms of the complete elliptic integrals
of the first and second kind.

In the numerical calculations we use only the kernels with $j=0$, 1, 2, and 3. Therefore, we
write down the explicit formulas only for these values of the total angular momentum
\begin{eqnarray}
\label{K0}
&K_{0}(k,q)=\frac{4}{k_{>}}{\rm K}(x),&\\
\label{K1}
&K_{1}(k,q)=\frac{4}{k_{>}}\frac{1}{x}\left[{\rm K}(x)-{\rm E}(x)\right],&\\
\label{K2}
&K_{2}(k,q)=\frac{4}{k_{>}}\frac{1}{3x^{2}}\left[(2+x^{2}){\rm K}(x)-2(1+x^{2}){\rm E}(x)\right],&\\
\label{K3}
&K_{3}(k,q)=\frac{4}{k_{>}}\frac{1}{15x^{3}}\left[(8+3x^{2}+4x^{4}){\rm K}(x)-(8+7x^{2}+8x^{4}){\rm E}(x)\right],&
\end{eqnarray}
where $k_{>}={\rm max}\{k,\, q\}$, $k_{<}={\rm min}\{k,\, q\}$ and $x=\frac{k_{<}}{k_{>}}$. By using formulas from Sec.~5.11 of Ref.~[\onlinecite{Gradstein}], we calculate the integrals
$\int\limits_{0}^{\Lambda}dq q K_{j}(k,q)$ with kernels (\ref{K0})-(\ref{K3}):
\begin{eqnarray}
\label{int0}
&{\int\limits_{0}^{\Lambda}dq q K_{0}(k,q)}=4\Lambda \mathrm{E}(\frac{k}{\Lambda}),&\\
\label{int1}
& {\int\limits_{0}^{\Lambda}dq q K_{1}(k, q)}=2k\left\{2G-1+{\int\limits_{k/\Lambda}^{1} \frac{\mathrm{E}(x)}{x} \, dx}
-\left[\mathrm{K}\left(\frac{k}{\Lambda}\right)-\mathrm{E}\left(\frac{k}{\Lambda}\right)\right]
\left(1-\frac{\Lambda^2}{k^2}\right)\right\},&\\
\label{int2}
&{\int\limits_{0}^{\Lambda}dq q K_{2}(k,q)}=\frac{4{\pi}k}{3}+\frac{8\Lambda^{3}}{9k^{2}}\left[\mathrm{K}\left(\frac{k}{\Lambda}
\right)-\mathrm{E}\left(\frac{k}{\Lambda}\right)\right]+\frac{4\Lambda}{9}\left[4\mathrm{K}
\left(\frac{k}{\Lambda}\right)-5\mathrm{E}\left(\frac{k}{\Lambda}\right)\right]- \frac{8k^{2}}{3\Lambda}\mathrm{K}\left(\frac{k}{\Lambda}
\right),&\\
\label{int3}
& {\int\limits_{0}^{\Lambda}dq q K_{3}(k, q)}=
\frac{8\Lambda^{4}}{15k^{3}}\left[\mathrm{K}\left(\frac{k}{\Lambda}\right)-\mathrm{E}
\left(\frac{k}{\Lambda}\right)\right]+ \frac{4\Lambda^{2}}{15k}\left[2\mathrm{K}\left(\frac{k}{\Lambda}\right)-3\mathrm{E}
\left(\frac{k}{\Lambda}\right)\right]-\frac{16k}{15}\left[\mathrm{K}\left(\frac{k}{\Lambda}\right)
-2\mathrm{E}\left(\frac{k}{\Lambda}\right)\right],&
\end{eqnarray}
where $G=0.915965$ is the Catalan constant. Since $K_{-j}(k,q)=K_{j}(k,q)$, we can use Eqs.~(\ref{int0})-(\ref{int3}) to calculate the critical charge for the total angular momentum from $j=-2$ to $j=2$.

\end{document}